# Spatially-resolved measurement of plasmon dispersion using Fourier-plane spectral imaging

Amir Ohad,[1,3] Katherine Akulov,[2,3] Eran Granot,[2] Uri Rossman,[1] Fernando Patolsky,[2] and Tal Schwartz[2,3,*]

[1]School of Physics, Raymond and Beverly Sackler Faculty of Exact Sciences, Tel Aviv University, Tel Aviv 69978, Israel

[2]School of Chemistry, Raymond and Beverly Sackler Faculty of Exact Sciences, Tel Aviv University, Tel Aviv 69978, Israel

[3]Tel Aviv University Center for Light-Matter Interaction, Tel Aviv University, Tel Aviv 69978, Israel

*Corresponding author: talschwartz@tau.ac.il

We show that Fourier-plane imaging in conjunction with the Kretschmann-Raether configuration can be used for measuring polariton dispersion with spatial discrimination of the sample, over the whole visible spectral range. We demonstrate the functionality of our design on several architectures, including plasmonic waveguides, and show that our new design enables the measurement of plasmonic dispersion curves of spatially-inhomogeneous structures with features as small as 3μm, in a single shot.

## 1. INTRODUCTION

Since the discovery of surface plasmon polaritons (SPPs) several techniques had been developed for their excitation and measurements [1,2]. The widely used Kretschmann-Raether prism-coupling configuration is common in most of the current applications of SPP's, primarily in sensing applications and thin film characterization [3]. These techniques rely on the high sensitivity of the surface plasmon resonance to the refractive index of the material in which the plasmons propagate. In order to measure the SPP resonance and its shift under material changes, two approaches are possible – either a broad-spectrum beam is reflected off the metallic surface at a fixed angle (spectral interrogation mode), or a monochromatic collimated beam is used and the reflected intensity is measured, while the angle of incidence is scanned (angular interrogation mode). It was shown in the past that by using Fourier-space (far-field) imaging in conjunction with the Kretschmann-Raether configuration, the two approaches can be unified to obtain the spectral and angular dependence of the plasmon resonance, in a single-shot measurement [4–8]. Alternatively, it was recently demonstrated [9] that similar Fourier-plane single-shot measurements can also be achieved by the use of a half-ball prism positioned on top of the sample. These techniques have the advantage that incident-angle scanning is not done mechanically (which requires a precisely-synchronized rotation of both the sample and the detector), but rather optically, taking advantage of the angular decomposition performed by lenses. Therefore, Fourier-plane imaging provides a convenient way to obtain the dispersion curves of surface plasmon over a broad spectral range and to characterize different plasmonic structures, such as periodic arrays or plasmonic waveguides. For sensing applications, it was demonstrated that measuring the reflection as a function of both wavelength and incident angle results in a better sensitivity [10,11]. This is because in such a measurement, in comparison to angular or spectral interrogation, more information is obtained and can be fitted numerically to extract refractive-index changes in the sample. Finally, the possibility to measure the plasmon dispersion in a simple and robust manner can also be used to extract the wavelength-dependent refractive index of an interrogated material, as used in SPP ellipsometry.

Here we demonstrate that the concept of Fourier-plane imaging can be further expanded to allow for spatially-resolved dispersion measurements in planar samples supporting surface plasmons. We show that by shaping the illumination beam and incorporating real-space imaging in the optical setup, pre- and post-selection of the interrogated area can be obtained. Therefore, our new optical design offers hybrid functionality and combines the high sensitivity of the broadband dispersion measurements with the imaging capabilities of SPR-imaging techniques. Our measurements show that the lateral detection resolution is high enough to acquire SPP dispersion curves in structures as small as 3μm.

## 2. EXPERIMENTAL SETUP

We designed the optical system as to enable the switching between two configurations, as shown in Fig. 1: a normal real-space imaging mode [Fig. 1(a)] and Fourier-plane imaging mode [Fig. 1(b)]. We use a fiber-bundle white light source (Olympus, LG-PS2) and pass the light beam through a 70 μm pinhole placed 15 cm after the exit of the fiber bundle. The diffracting beam is passed through a linear polarizer, collimated with a 250 mm lens (L1), and then sent into a right-angle prism (BK7, WZW Optics, RP10) at normal incidence. The sample is attached to the hypotenuse of the prism with immersion oil (refractive index 1.516) and the prism is mounted on a translation stage, allowing the lateral displacement of the sample together with the prism. The diameter of the collimated beam as it enters the prism is approximately 10 mm and after being reflected from the sample it is projected by the lenses L2 (focal length 50mm) and L3 (focal length 75mm) onto the entrance slit of an imaging spectrometer (Isoplane SCT 320, Princeton Instruments) equipped with a CCD detector (Pixis 1024, Princeton Instruments) and a 50 lines/mm grating. L2 is positioned such that its front focal plane coincides with the sample and the spectrometer entrance slit (set to a width of 20 μm) is located at the back focal plane of L3. Therefore, these two lenses form an imaging system with a 1.5X magnification and image the sample onto the entrance plane of the spectrometer. In the imaging mode we set the angle of the spectrometer grating to 0 degrees, which projects the resulting intermediate image from the plane of the spectrometer entrance slit to the CCD. Fig. 1(c) shows an image of a 1951 USAF resolution target, as recorded in the real-space imaging mode with the spectrometer slit removed. As can be seen, the spatial resolution of the imaging is sufficient for imaging features smaller than 8 μm.

In order to switch to the Fourier-plane imaging mode [Fig. 1(b)], we first insert a cylindrical lens (C1, focal length 50 mm) before the prism, which focuses the beam into a narrow horizontal strip of 14 μm FWHM, exactly on the sample, as shown in Fig. 1(d). By this, we achieve two goals – first, as also used in Refs. [4–8], the sample is simultaneously probed by the many plane-waves making up the converging beam and having an angular span of 10 degrees (in air). Using Snell's law, this is translated to a range of incident angle of 41°-48° with respect to the sample. Second, by probing the sample with a focused beam, we obtain spatial localization of 20 μm of the interrogated area in the vertical dimension (taking into account the 45° tilt of the sample). The spectrometer slit, set to a 20 μm width, provides us with lateral localization of 13.3 μm (due to the 1.5X magnification of the imaging system), as it post-selects only a small section out of the illuminated area of the sample. Therefore, by limiting the interrogated area both in the vertical and horizontal directions, we can obtain a local measurement of the dispersion. By laterally translating the sample (together with the prism) our setup can be used for spatially-resolved measurements of inhomogeneous samples, as we demonstrate below (Section 4). To perform the optical Fourier transform, a negative cylindrical lens (C2, focal length −30 mm) is placed after the prism, such that a plane-wave entering C2 is transformed into a virtual image of a horizontal line, located behind C2. This virtual horizontal line is displaced from the optical axis (in the vertical direction) by a distance which depends on the entrance angle of the plane-wave into C2. We adjust the exact position of C2 by bringing this virtual image to coincide with the plane of the sample, such that the thin horizontal line is refocused at the slit by the L2-L3 lens pair, as seen in Fig. 1(e). In that way, the spectrometer images the Fourier plane of C2, which corresponds to the angular distribution along the vertical (y) direction. As the angular span of the light-beams around the optical axis is small (±4 degrees), we can use the paraxial approximation to relate the vertical position on the CCD with the propagation angle ($\theta_a$) of a plane-wave entering C2 by

$$y = y_0 + 1.5 \times f_{C2} sin\theta_a \qquad (1)$$

where $y_0$ is correspond to the intersection between the optical axis and the CCD, and we have taken into account the 1.5X magnification of the imaging lenses. Furthermore, denoting the incident angle of the light inside the prism (with respect to the normal to the sample) by $\theta_i$ and using Snell's law, this angle is given by the relation

$$\theta_i = 45° + \sin^{-1}\left[\frac{1}{n_p}\sin\theta_a\right] \qquad (2)$$

with $n_p$=1.5 being the refractive index of the BK7 prism. Note that the width of the focused beam in the Fourier-plane [see Fig. 1(e) is 9 μm FWHM, meaning that the angular resolution of our measurement is 0.7 arcminute. Finally, the spectrometer disperses the various wavelengths contained in the white light beam along the horizontal direction (x) and by this we obtain a full mapping the angle-resolved reflection spectrum of the sample, in a similar manner to Ref. [8]. Note that for all the lenses besides C2 we use achromatic doublets in order to avoid chromatic aberrations. In addition, the cylindrical lenses C1 and C2 are mounted on a flip-mount to allow a convenient and reproducible removal and insertion of the lenses to switch between the real-space and Fourier-space imaging modes. This convenient switching between different imaging modes provides us with the ability to accurately select the interrogated area within the sample, as we demonstrate below (see Section 4).

Compared to all previous works, the main strength of our configuration is that it allows the acquisition of the plasmonic dispersion relation with spatial discrimination and a resolution of several microns. This advantage stems from the combination of an imaging system (formed by L2 and L3) and 1-D Fourier imaging by C2, along with the pre-selection of the illuminated area by C1 and the post-selection of the measured area by the spectrometer slit. Moreover, the use of a negative lens (C2) to perform the optical Fourier transform ensures that the length of the optical path is minimal, resulting in a higher numerical aperture for given lens sizes. With the specific choice of parameters used, our setup provides us with the ability to measure the dispersion over an area as small as 13.3×20 μm². In principle, the interrogated area can be made even smaller, by using stronger lens for C1, or higher magnification in the imaging subsystem. Alternatively, reducing the pinhole size, or the width of the spectrometer entrance slit, will also result in a better spatial discrimination, but this will come at the expense of intensity and signal-to-noise ratio.

In the Kretschmann configuration the resonance angle for each wavelength is determined by momentum conservation between the SPP propagating along the film and the plane-wave impinging on the sample through the prism. This condition is expressed by

$$k_{spp} = k_0 n_p \sin\theta_i \tag{3}$$

where $k_0 = 2\pi/\lambda = \omega/c$ is the wavenumber in vacuum, $\lambda$ the wavelength of light, $\omega$ is the angular frequency and c the speed of light. Using this expression, the dispersion relation $\omega(k_{spp})$ can be extracted from the measurement. Alternatively, and as will be used below, the SPP dispersion relation can be presented in terms of the wavelength-dependent effective index for the plasmon propagating along the interface, which is defined by

$$n_{eff} = \frac{k_{spp}}{k_0} = n_p \sin\theta_i \ . \tag{4}$$

Using Eqs. (1)-(4), one finds that in our setup the effective index of refraction can be related to the vertical position on the CCD using

$$n_{eff} = n_p \sin\left[45° + \sin^{-1}\left(\frac{1}{n_p} \times \frac{y-y_0}{1.5 f_{C2}}\right)\right] . \tag{5}$$

In order to determine the value of $y_0$, we remove the focusing lens C1, which results in a single plane-wave being reflected from the sample at 45°, propagating parallel to the optical axis and finally focused at the Fourier plane [as shown in Fig. 1(e)]. This calibration procedure was performed before each set of measurements in order to compensate for small misalignments of the system. Spectral calibration was performed using a Mercury-vapor discharge lamp.

## 3. PLASMON DISPERSION ON PLANAR FILMS

First, we tested the performance of our setup on a simple planar Ag film. The sample was fabricated by sputtering a 50±2 nm layer of Ag on a glass substrate. In order to obtain normalized reflection maps, we initially measured the sample with TE-polarized light, for which no plasmons are excited and the reflectivity of the Ag film depends weakly on either the wavelength or the angle of incidence. Figure 2(a) shows the resulting reflected intensity map as a function of wavelength (horizontal axis) and position $y$ (vertical axis). We then rotated the polarizer and repeated the measurement with TM polarization [Fig. 2(b)]. By taking the ratio between these two images we obtain a normalized reflectivity map, and using Eq. (5) we transform the resulting image to obtain the plasmonic dispersion $n_{eff}(\lambda)$, which is clearly visible as a dark notch in Fig. 2(c). We obtained good S/N ratios throughout the spectral range of 420-850nm and an effective index range of 1-1.11. This range is limited by the numerical aperture of the system and can be increased by the use of larger lenses. The white solid line superimposed on the image corresponds to the theoretical dispersion curve of SPP propagating on the Ag-air interface, which was calculated using transfer-matrix simulations (with the dielectric function of Ag measured by ellipsometry). Our measurement exhibits excellent agreement with the numerical simulations throughout the whole spectral range, confirming the accuracy of our measurement technique.

As another example for the capabilities of the Fourier-plane imaging technique, we used our setup to measure the dispersion of exciton-plasmon polaritons [12–15], formed by the strong coupling of surface plasmons to thin films of organic dye molecules. For this purpose, we prepared a sample composed of a 50 nm-thick Ag film on which we spun-coated a thin layer of a transparent PVA polymer (polyvinyl alcohol), doped with TDBC molecules (Few Chemicals, S0046). These molecules form J-aggregates which have a sharp absorption peak at 590 nm and they were previously used for studying strong light-matter coupling in both plasmonic and microcavity system [12,16,17]. Under strong coupling conditions, when the dispersion curve of the electromagnetic modes crosses the molecular absorption, the dispersion splits into two separate curves known as the upper and lower polariton branches. This behavior is clearly observed in Fig. 2(d), which shows the dispersion measured in a similar manner as for the bare silver film. Around the exciton-plasmon resonance, we find a splitting of 130 meV, which is within the typical range of values measured previously. Note that, as expected, the plasmonic dispersion is pushed toward higher effective indices, due to the presence of the high-index PVA layer on top of the Ag film. Once again, we compared the experimental measurement to numerical simulations, which are represented by the solid lines in Fig. 2(d). In these simulations, the thickness of the PVA layer (taken with a refractive index of 1.5) was adjusted by fitting the calculated dispersion to the measured one. Using this fitting procedure we found that the PVA/TDBC layer has a thickness of 20 nm and an absorbance of 0.08 at the exciton wavelength (590 nm).

## 4. DISPERSION OF PLASMONIC WAVEGUIDES

In the two former examples the measured structures were spatially-homogeneous. However, as we designed our measurement system to have spatial discrimination capabilities, we can use it to perform local dispersion measurements on inhomogeneous samples. In order to demonstrate this unique functionality, we fabricated a series of several plasmonic metal-strip waveguides [2,18], which support propagation of surface plasmons along the waveguides. Similar structures were previously measured using leakage-radiation microscopy [19,20], and here we show that such structures can also be measured using our method, which is compatible with standard SPR-sensing configurations. Fig. 3(a) shows a microscope image of the array, which consists of 200 μm-long metal-strip waveguides with varying width ($w$=50, 10, 3, 2.5, 2 and 1.5 μm). In addition, we fabricated on the same sample a square Ag patch of 200×200 μm$^2$, emulating a homogeneous metal film. This sample was fabricated using e-beam lithography, followed by evaporation of $h$=50 nm-thick Ag layer. Finally, in order to protect the Ag film from oxidation, we deposited a thin layer of Al$_2$O$_3$ using atomic-layer deposition (ALD). A sketch of the resulting waveguide structure is shown in Fig. 3(b). In order to measure the different waveguides, we switched our setup to imaging mode by removing lenses C1 and C2 and aligned the sample such that a particular waveguide is located at the center of the field-of-view, with its image coinciding with the spectrometer entrance slit. The system was then switched back to the Fourier-imaging mode in order to measure the dispersion curve of the individual waveguides. The reflection reference used for the waveguide-array sample was recorded by measuring an area in-between the waveguides (glass covered Al$_2$O$_3$), in TM polarization.

In Fig. 4(a)-(d) we plot the results of the Fourier-plane imaging measurement, taken for the 200 μm square patch (a) and for waveguides with a width of 50 (b), 10 (c) and 3 μm (d). The dispersion curve of the SPP is clearly seen for all of the waveguides, which demonstrates our ability to measure the dispersion locally, down to structures as small as 3 μm. For the narrower waveguides, the images were too blurred to identify the plasmonic dispersion. In order to validate the results in Fig. 4, we compared them to our numerical simulations. Note that for waveguide widths of several microns or more, the plasmonic dispersion barely changes as a result of the lateral confinement [21], making it practically identical to the dispersion of plasmons propagating on a uniform surface. Moreover, as the reflected field is uniform across the width of the waveguides, it cannot couple to high-order plasmonic modes [21] and only a single branch is seen in the measurements. This permits us to use the same transfer matrix simulations as in Section 3, with the addition of the thin Al$_2$O$_3$ layer on top of the Ag film. Moreover, by fitting the results of the simulations to the measured dispersion curves, we are able to extract the thickness of the Al$_2$O$_3$ layer. The dashed line in Fig. 4(a) corresponds to the SPP dispersion calculated for an Ag-air interface, as in Fig. 2(c), whereas the solid lines are the dispersion curves calculated for Ag layer covered with a layer of 7.5 and 10 nm thick Al$_2$O$_3$. As seen in Fig. 4(a), the measured dispersion is well situated between these two lines over the whole spectral range between 500 and 850 nm, resulting in a thickness sensing resolution better than 1.5 nm. As previously demonstrated, by using more sophisticated spectral fitting methods [11,22] a further increase in accuracy can be obtained when extracting optical properties from complete dispersion curves. For the waveguide structures [Figs. 4(b)-(d)], we plot the same two simulated curves and find that the measured SPP dispersion curves still lie within the same error margins, meaning that the sensing capabilities are maintained even for the measurement over the smallest structure of 3 μm. Therefore, our results indicate that the Fourier-plane imaging setup with our new design can be used for high-resolution sensing applications, with a spatial resolution of several microns.

It is interesting to note that in Figs. 4(a) and 4(b), corresponding to relatively large widths, the plasmon dispersion appears as a dark notch in the reflectivity map, which is normal for Kretschmann-Raether measurements. However, for waveguide widths of 10 μm and less, the same dispersion curve appears as a signal higher than the background, even though it has the same shape. We suspect that this is a result of diffraction effects, which arise when the width of the waveguide is comparable to the spatial resolution limit of the imaging system. For such small structures, part of the light will not be collected and re-focused on the spectrometer entrance slit, resulting in a reflection baseline lower than 100%, as for a planar, homogeneous film However, waves which fall on the plasmon dispersion curve are coupled to the waveguide, and therefore such scattering losses are weaker.

In conclusion, we presented a new optical design which is based on the Kretschmann-Raether configuration and incorporates both Fourier-plane spectral imaging and real-space imaging as a convenient platform for performing spatially-resolved, single-shot measurements of plasmonic dispersion curves. We demonstrated the measurements of SPP dispersion on bare planar Ag films and for SPP waves strongly-coupled to molecular films, over a broad spectral range of 420-850nm. Taking advantage of the spatial-interrogation capabilities, we measured the SPP dispersion curves of metal-strip waveguides and demonstrated that our method allows local measurement of waveguides which are as narrow as 3 μm. Like previous Fourier-space imaging methods used for dispersion measurements [4–9], our method can be used for broadband, high-resolution and real-time sensing applications. However, in addition, it permits spatial discrimination of several microns and mapping of spatially-inhomogeneous structures, by lateral scanning over the interrogated area. As such, our new method stands in the gap between conventional SPR sensors, which provide high sensing resolution, and plasmon imaging techniques, such as leakage-radiation microscopy [19,20,23,24] or scanning near-field microscopy, which provide high spatial resolution, but require elaborate optical setups and precise alignment. Finally, we stress that in our system, as with other prism-based configurations, the angular span of the excitation beam is relatively narrow and matches the momentum range of the plasmonic modes. In comparison to leakage-radiation microscopy, where the angular spread of the focused beam is very broad (corresponding to the full aperture of the focusing objective), our method provides maximal coupling of light into surface plasmons, and therefore higher signal-to-noise ratio and higher sensitivity.

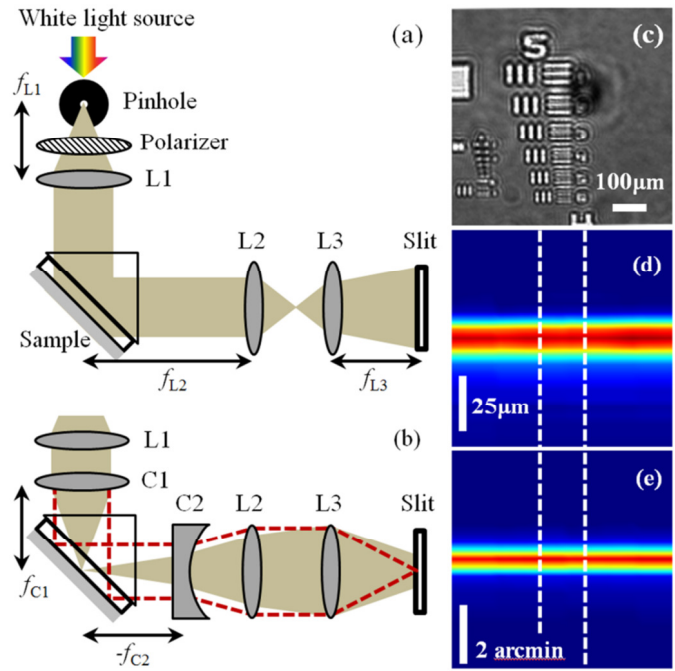

Fig. 1. Schematic sketch of the measurement system in (a) real-space imaging mode and (b) in Fourier-plane imaging mode. The dashed red lines illustrate the evolution of a single plane-wave entering the prism and propagating through the setup in Fourier-plane imaging mode. (c) A subsection of a resolution target, as imaged through the optical system in the real-space imaging mode. (d) Real-space intensity distribution at the sample plane after focusing by L1. (e) Intensity distribution in Fourier-plane imaging mode for an incoming collimated beam. The dashed vertical lines in (d) and (e) mark the edges of the spectrometer entrance slit.

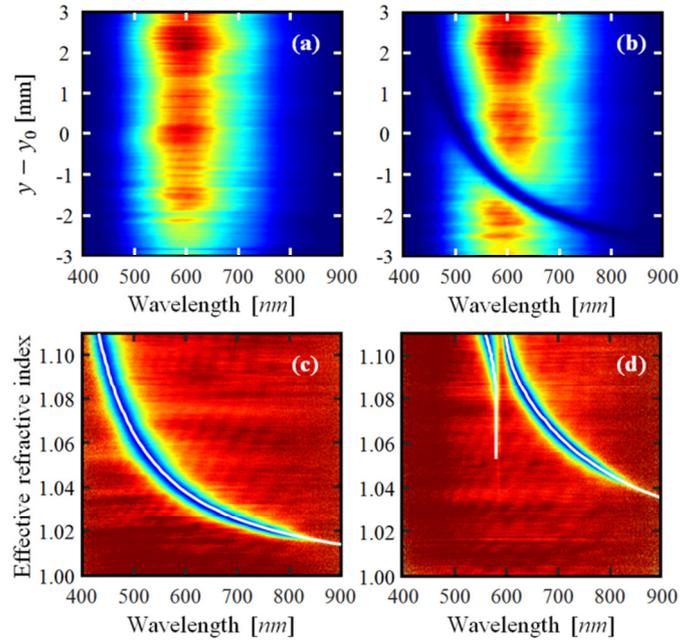

Fig. 2. Reflected intensity maps recorded for an Ag film for TE (a) and TM (b) polarization. (c) Dispersion diagram of SPP on Ag film, measured by Fourier-plane spectral imaging. The solid line shows the SPP dispersion curve calculated by T-matrix method. (d) Measured dispersion curve for an Ag film covered with a PVA/TDBC layer, exhibiting normal-mode splitting as a result of strong coupling between SPP's and molecular excitons. The solid lines show the dispersion curve calculated by T-matrix method with the PVA thickness and the molecular absorption as free parameters.

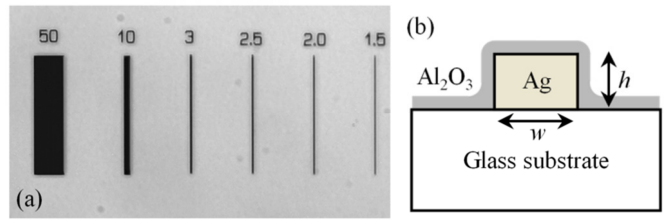

Fig. 3. (a) An optical microscope image of the fabricated metal-strip waveguides. The numbers indicate the width of the individual waveguides. (b) A schematic sketch of a waveguide cross-section.

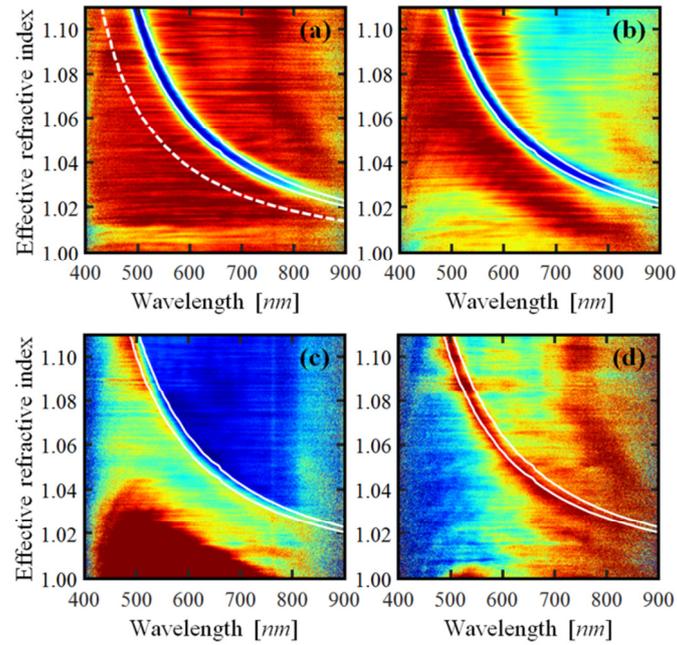

Fig. 4. SPP dispersion measured by Fourier-plane spectral imaging for a 200μm×200μm Ag patch covered with Al$_2$O$_3$ (a) and for metal-strip waveguides of widths $w = 50\mu m$ (b), $w = 10\mu m$ (c) and $w = 3\mu m$ (d). The solid lines correspond to the simulated dispersion with an Al$_2$O$_3$ layer in the range of 7.5-10 nm and the dashed line in (a) corresponds to the dispersion curve calculated for a bare Ag film.